\documentclass[aps,pra,preprint,showkeys,letterpaper]{revtex4}
\usepackage{amsmath,amssymb}
\usepackage{bm}
\usepackage{graphicx}
\usepackage[colorlinks=true,allcolors=blue]{hyperref}

\begin{document}
	
\title{Bistability and Exact Reflectionless States in Nonlinear Scattering of a Bose--Einstein Condensate}

\author{Feilong Wang}
\affiliation{College of Physics and Electronic Engineering \& Key Laboratory of Laser Technology and Optoelectronic Functional Materials of Hainan Province, Hainan Normal University, Haikou 571158, China}

\author{Jinlin Fan}
\affiliation{College of Physics and Electronic Engineering \& Key Laboratory of Laser Technology and Optoelectronic Functional Materials of Hainan Province, Hainan Normal University, Haikou 571158, China}

\author{Ruolin Chai}
\affiliation{Center for Theoretical Physics \& School of Physics and Optoelectronic Engineering, Hainan University, Haikou 570228, China}

\author{Zhibin Zhao}
\email[Email: ]{zhaozhibin@hainnu.edu.cn}
\affiliation{College of Physics and Electronic Engineering \& Key Laboratory of Laser Technology and Optoelectronic Functional Materials of Hainan Province, Hainan Normal University, Haikou 571158, China}

\author{Qiongtao Xie}
\email[Corresponding author: ]{xieqiongtao@hainnu.edu.cn; xieqiongtao@126.com}
\affiliation{College of Physics and Electronic Engineering \& Key Laboratory of Laser Technology and Optoelectronic Functional Materials of Hainan Province, Hainan Normal University, Haikou 571158, China}

\begin{abstract}
We investigate the mean-field scattering dynamics of a quasi-one-dimensional Bose--Einstein condensate interacting with a Rosen--Morse potential. For specific potential and nonlinearity parameters, we derive analytically exact, degenerate scattering states (doubly or triply degenerate) exhibiting perfect transmission. Using the Bogoliubov--de Gennes approach, we analyze the stability of these reflectionless degenerate states, demonstrating that only one solution within each degenerate manifold is dynamically stable. Furthermore, we study a configuration with spatially localized nonlinearity, identifying an exact reflectionless state under specific conditions. Numerical analysis shows that this state marks the system's transition from monostability to bistability as the incident wave amplitude increases. Our work establishes an analytic framework for these multistable transmission phenomena, directly relevant to coherent matter-wave transport in ultracold atomic systems and optical propagation in engineered photonic lattices.
\end{abstract}

	
\maketitle

\section{Introduction}
Quantum scattering of a single particle from potential wells or barriers represents a fundamental wave phenomenon, with investigations tracing back to the early days of quantum mechanics~\cite{Epstein1930,Eckart1930}. Owing to their long coherence times and high parameter controllability, systems of Bose--Einstein condensates (BECs) provide a powerful platform for exploring the influence of many-body interactions on scattering dynamics. Experimentally, quantum reflection of BECs from various silicon surfaces has been studied~\cite{Ketterle2004,Pritchard2006}. While quantum reflection for individual atoms typically exhibits a monotonic decrease with increasing incident velocity, the presence of atomic interactions introduces novel effects. Specifically, interaction-induced suppression of quantum reflection at low velocities has been observed~\cite{Ketterle2004,Pritchard2006}. Theoretically, this anomalous low-velocity reflection has been attributed to the formation of solitons and vortex rings mediated by interatomic interactions~\cite{Sheard2005}. 

Bistability—the coexistence of two stable states and, more generally, multistability—is a hallmark of nonlinear systems~\cite{Abraham1982,Gibbs1985}. It manifests across diverse physical platforms, including optical cavities~\cite{Girvin2006}, cavity magnonics~\cite{You2018}, and spinor-polariton condensates~\cite{Lagoudakis2018}. In the Gross--Pitaevskii mean-field description, interatomic interactions endow a BEC with an effective nonlinearity. Experimental observations of bistability are well established in several BEC settings, including spinor BECs~\cite{Oberthaler2010}, driven--dissipative BECs in one-dimensional optical lattices~\cite{Ott2016}, BECs in ring cavities~\cite{Goldwin2018}, and spin-orbit-coupled BECs under Raman-quench protocols~\cite{Chen2020}.

For quantum scattering, the mean-field nonlinearity strongly reshapes the reflection and transmission response, departing from the single-particle picture~\cite{Schlagheck2005a,Schlagheck2005b,Korsch2005,Korsch2006,Korsch2008,Strong2012}. A key consequence is a multivalued transmission coefficient in certain parameter regimes—an S-shaped incident-transmission curve with hysteresis—signaling bistability~\cite{Schlagheck2005a,Schlagheck2005b,Korsch2005,Korsch2006,Strong2012}. Related lines of work have also examined scattering of matter-wave solitons and quantum droplets, revealing rich nonlinear phenomena across diverse configurations~\cite{Kivshar2002,Pavloff2005,Abdullaev2006,Frantzeskakis2006,Lee2006,Adams2009,Luo2023}.

Despite extensive studies of quantum scattering with BECs, closed-form expressions for reflection and transmission coefficients are available only in special cases~\cite{Korsch2006,Korsch2008,Strong2012}. For the Rosen--Morse (RM) potential~\cite{Epstein1930,Eckart1930}, an exact solution yielding completely reflectionless transmission at the first transmission resonance has been derived, together with an analytic expression for the associated nonlinear resonance shift~\cite{Ishkhanyan2010,Ishkhanyan2012}. Nevertheless, such exact reflectionless solutions remain scarce beyond these conditions. Critically, the dynamical stability of these exact solutions has not been investigated systematically.

In this work, we analyze the nonlinear scattering of a quasi-one-dimensional (1D) BEC from a Rosen--Morse (RM) potential within the mean-field approximation. Utilizing the trial solution approach, we identify both two-fold and three-fold degenerate reflectionless states at specific values of the potential and nonlinearity parameters. Bogoliubov--de Gennes (BdG) stability analysis demonstrates that only one exact reflectionless state is dynamically stable. Additionally, we investigate the case of a localized nonlinearity, finding that the BEC supports an exact reflectionless state under specific parameter conditions. Our simulations reveal that this reflectionless state connects closely to the mono-to-bistable transition triggered by increasing the incident wave amplitude. Our analytical results explicitly confirm multistable transmission phenomena. Given the established feasibility of realizing RM potentials in ultracold atomic systems ~\cite{Juzeliunas2006} and engineered photonic lattices ~\cite{Szameit2011}, these findings are poised for direct application in such systems.

\section{Theoretical model and exact reflectionless states}
\label{sec:RM-multiple-exact}
\subsection{Theoretical model}
We consider a quasi-1D BEC trapped in the well-known RM potential. Within the mean-field framework, the macroscopic condensate wavefunction $\psi(x,t)$ satisfies the dimensionless Gross--Pitaevskii equation (GPE)~\cite{Dalfovo1999,Gross1961,Pitaevskii1961}
\begin{equation}
	\mathrm{i}\,\frac{\partial \psi(x,t)}{\partial t}
	= -\frac{1}{2}\,\frac{\partial^2 \psi(x,t)}{\partial x^2}
	+ V(x)\,\psi(x,t) + g\,|\psi(x,t)|^2\psi(x,t),
	\label{eq:GPE_dimless}
\end{equation}
where $V(x)$ is the RM potential~\cite{Rosen1932,Moses1956}
\begin{equation}
	V(x) = -V_0\,\operatorname{sech}^2(x),\qquad V_0>0. 
\end{equation}
Here the spatial coordinate $x$, time $t$, and potential depth $V_0$ are measured in $1/\alpha$, $m/(\hbar\alpha^2)$, and $\hbar^2\alpha^2/m$, respectively, where $m$ denotes the atomic mass and $\alpha$ is the RM width parameter. The wavefunction $\psi(x,t)$ is normalized by $\sqrt{n_0}$, the averaged BEC density. The nonlinearity parameter $g$ is defined as
$g = m n_{0} g_{1\mathrm{D}} / (\hbar^{2}\alpha^{2})$,
with $g_{1\mathrm{D}}$ the 1D interaction parameter.

To study stationary scattering states, we substitute $\psi(x,t)=\phi(x)\,e^{-\mathrm{i}\mu t}$ into Eq.~\eqref{eq:GPE_dimless}, yielding
\begin{equation}
	-\frac{1}{2}\,\frac{d^2\phi(x)}{dx^2} + V(x)\,\phi(x) + g\,|\phi(x)|^2\phi(x) = \mu\,\phi(x).
	\label{eq:stationary}
\end{equation}
Here, $\mu$ is the chemical potential. In the following, we consider the scattering states with perfect transmission \cite{Schlagheck2005a,Schlagheck2005b}, and then the wavefunction $\phi(x)$ displays the asymptotic behavior, $\phi(x) \sim e^{ \mathrm{i} k x}$ as $x \rightarrow-\infty$ and $\phi(x) \sim \tilde{t} e^{ \mathrm{i} k x}$ as $x \rightarrow+\infty$, where $\tilde{t}$ is the complex transmission amplitude. Probability-current conservation requires $|\tilde{t}|=1$, allowing the transmission amplitude to be written as a pure phase factor $\tilde{t}=e^{ \mathrm{i} \theta}$, where $\theta$ is the phase shift. The asymptotic decay $V(x)\to0$ as $|x|\to\infty$ imposes the dispersion relation
\begin{equation}
	\mu = g + \frac{k^2}{2},\qquad
	k=\sqrt{2(\mu-g)} > 0,
	\label{eq:dispersion}
\end{equation}
indicating that propagating solutions exist only for $\mu>g$; otherwise, $k$ becomes imaginary, leading to evanescent tails. Crucially, Eq.~\eqref{eq:dispersion} reveals that the far-field wavenumber $k$ depends explicitly on the nonlinearity $g$, distinguishing the nonlinear case from its linear counterpart. In the linear limit $g=0$, the RM potential exhibits reflectionless transmission for the discrete well depths~\cite{Rosen1932,Moses1956}
\begin{equation}
	V_0=\frac{n(n+1)}{2},\qquad n=1,2,\dots.
\end{equation}
Notably, such reflectionless states persist for selected parameters in the nonlinear regime $(g\neq 0)$~\cite{Ishkhanyan2010}.

\subsection{Isolated exact reflectionless state}
In prior work, an isolated exact reflectionless state was identified near the first resonance ($n=1$). By adjusting the well depth through nonlinearity to the value~\cite{Ishkhanyan2010,Ishkhanyan2012}
\begin{equation}
	V_0 = 1 - \frac{g}{1 + 2(\mu - g)}
	= 1 - \frac{g}{1 + k^2},
	\label{eq:V0-nl}
\end{equation}
Eq.~\eqref{eq:stationary} admits the exact isolated solution:
\begin{equation}
	\phi_{\mathrm{I}}(x)
	= \frac{\mathrm{i} k - \tanh x}{1 + \mathrm{i} k}\,\mathrm{e}^{\mathrm{i} k x}
	= \frac{\mathrm{e}^{\mathrm{i} k x}}{\mathrm{e}^{-x} + \mathrm{e}^{x}} \left( \mathrm{e}^{-x} + \tilde{t}_{\mathrm{I}} \mathrm{e}^{x} \right),
	\label{eq:phi-FamilyI}
\end{equation}
with a transmission amplitude 
\begin{equation}
    \tilde{t}_{\mathrm{I}} = \frac{\mathrm{i} k - 1}{1 + \mathrm{i} k} = e^{\mathrm{i}\theta_{\mathrm{I}}},
    \label{eq:t0-exact}
\end{equation}
which satisfies the perfect transmission condition $\left|\tilde{t}_{ \mathrm{I} }\right|=1$. We emphasize that this exact reflectionless state holds for arbitrary interaction strength $g$ provided $\mu>g$. Its existence at $g=0$ (i.e., the linear limit) confirms it originates from a linear counterpart.

Motivated by the mathematical structure of this solution, we introduce a new two-term trial solution that enables the systematic construction of a complete family of reflectionless solutions, as demonstrated in what follows.

\subsection{Degenerate exact reflectionless states}
Following the form of the isolated exact reflectionless state in Eq.~\eqref{eq:phi-FamilyI}, we use the following ansatz 
\begin{equation}
	\phi(x)=\frac{e^{\mathrm{i} k x}}{\sqrt{e^{-x} + e^{x}}} \Big( e^{-x/2} + \tilde{t} \, e^{x/2} \Big).
	\label{eq:familyII}
\end{equation}
Substituting Eq.~\eqref{eq:familyII} into the stationary equation 
(see Appendix~\ref{app:family-II} for details) yields the algebraic constraints

\begin{equation}
	V_0=\frac{3}{8}, \qquad \mu=\frac{3}{2}g-\frac{1}{8}, \qquad k = \sqrt{2(\mu - g)} = \frac{1}{2} \sqrt{4g - 1}.
\end{equation}
Remarkably, under these conditions, we find a pair of degenerate reflectionless solutions, $\phi_{\mathrm{II},\pm}(x)$, which take the explicit form
\begin{equation}
	\phi_{\mathrm{II},\pm}(x)=\frac{e^{\mathrm{i} k x}}{\sqrt{e^{-x}+e^{x}}}
	\left(
	e^{-x/2} + \tilde{t}_{\mathrm{II},\pm}\,e^{x/2}
	\right).
	\label{eq:phi-FamilyII}
\end{equation}
The transmission amplitudes for these states are
\begin{equation}
	\tilde{t}_{\mathrm{II},\pm} = \pm \frac{2 \sqrt{g}}{1 + 2 \mathrm{i} k} = e^{\mathrm{i}\theta_{\mathrm{II},\pm}},
	\qquad
	|\tilde{t}_{\mathrm{II},\pm}|=1.
	\label{eq:Tpm}
\end{equation}
Hence the branches $\phi_{\mathrm{II},+}$ and $\phi_{\mathrm{II},-}$ constitute a two-fold degenerate family, distinguished by a constant phase of $\pi$: $\theta_{\mathrm{II},-}=\theta_{\mathrm{II},+}+\pi$.

We emphasize that these degenerate reflectionless states exist only for $g>1/4$, indicating that they have no linear counterparts. Additional families of degenerate reflectionless states derived from distinct ansatze are presented in Appendix~\ref{app:family-III}.

\begin{figure}[t]
	\centering
	\includegraphics{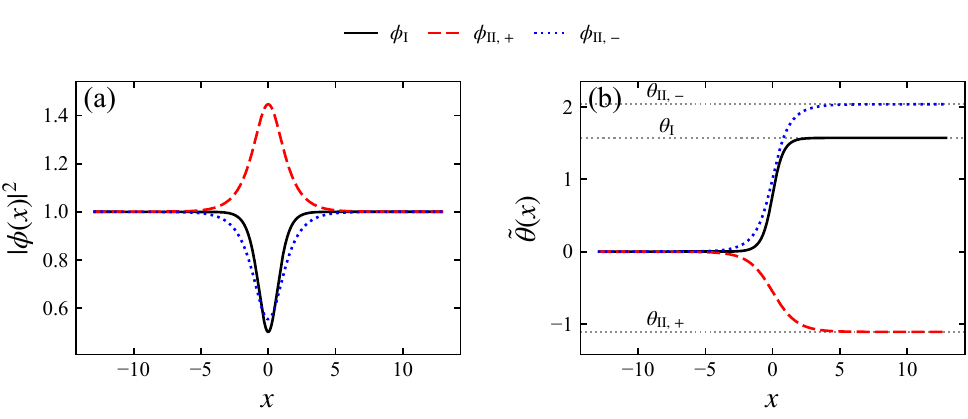}
	\caption{Three-fold degenerate reflectionless stationary states for the parameter values $V_0=3/8, \mu=7/4, g=5/4$, and $k=1$ (dimensionless units).
		(a) Densities $|\phi_{\mathrm{I}}(x)|^2$ and $|\phi_{\mathrm{II},\pm}(x)|^2$.
		(b) Phase $\tilde{\theta}(x)=\arg \left[\phi(x) e^{-\mathrm{i} k x}\right]$. The horizontal dotted lines show the analytically predicted asymptotic phase shifts ($\theta_{\mathrm{I}},\theta_{\mathrm{II},\pm}$).}
	\label{fig:fig1}
\end{figure}

\subsection{Three-fold degenerate reflectionless states}
Significantly, we find that the degenerate reflectionless states $\phi_{\mathrm{II},\pm}(x)$ [Eq.~\eqref{eq:phi-FamilyII}] and the isolated reflectionless state $\phi_{\mathrm{I}}(x)$ [Eq.~\eqref{eq:phi-FamilyI}] exist for a common set of parameters $\{V_0,\mu,g,k\}$. This results in a set of three-fold degenerate reflectionless states at the specific values
\begin{equation}
	V_0=\frac{3}{8},\qquad \mu=\frac{7}{4},\qquad g=\frac{5}{4},\qquad k=1.
\end{equation}
At these parameters, the transmission amplitudes $\tilde{t}_{\mathrm{I}}$ and $\tilde{t}_{\mathrm{II},\pm}$ given previously [Eqs.~\eqref{eq:t0-exact} and \eqref{eq:Tpm}] possess unit magnitude $|\tilde{t}_{\mathrm{I}}|=|\tilde{t}_{\mathrm{II},\pm}|=1$ but carry different phases. Thus, while all three branches exhibit perfect transmission, they represent distinct states characterized by unique far-field transmission phases, as illustrated in Fig.~\ref{fig:fig1}.

Figure~\ref{fig:fig1}(a) shows the densities for all states approaching unity as $|x|\to\infty$, consistent with unit transmission. However, their spatial profiles differ substantially: $\phi_{\mathrm{II},+}(x)$ exhibits a localized bright hump on a nonvanishing background, resembling an antidark (or dark-like bright) profile~\cite{Kivshar1996}. In contrast, $\phi_{\mathrm{I}}(x)$ and $\phi_{\mathrm{II},-}(x)$ display gray-soliton-like density notches~\cite{Kivshar1998}.

The phase evolution shown in Fig.~\ref{fig:fig1}(b) starts from $0$ as $x\to-\infty$ and asymptotically approaches $\theta_{\mathrm{I, II, \pm}} = \arg(\tilde{t}_{\mathrm{I, II, \pm}})$ as $x\to+\infty$. For $k=1$, these phases are $\theta_{\mathrm{I}}=\pi/2$, $\theta_{\mathrm{II},+}=-\arctan(2)$, and $\theta_{\mathrm{II},-}=\pi-\arctan(2)$, matching the theoretical limits ($\arg(\tilde{t})$), indicated by the horizontal reference lines. Therefore, despite sharing identical far-field properties (unit density and transmission magnitude), the three states are unambiguously distinguished by their internal density and phase structures.

\section{Stability analysis}
Exact solutions are experimentally relevant only if they are dynamically stable over the observation timescale. We assess stability by linearizing the time-dependent GPE around a stationary state $\phi(x)$ of Eq.~\eqref{eq:stationary}. Using the standard BdG approach, we express the perturbed wavefunction as \cite{Frantzeskakis2010,Bogoliubov1947}
\begin{equation}
	\psi(x,t)=\Big[\phi(x)+u(x)\,\mathrm{e}^{\varepsilon t}+v^*(x)\,\mathrm{e}^{\varepsilon^* t}\Big]\mathrm{e}^{-\mathrm{i}\mu t}.
\end{equation}
The linearization yields the non-Hermitian eigenproblem
\begin{equation}
	\mathcal L \binom{u}{v}=\mathrm{i}\,\varepsilon\binom{u}{v},
\end{equation}
with
\begin{equation}
	\mathcal L=
	\begin{pmatrix}
		\mathcal H & g\,\phi^2\\
		-\,g\,\phi^{*2} & -\mathcal H
	\end{pmatrix},
	\qquad
	\mathcal H=-\frac12\,\frac{\mathrm{d}^2}{\mathrm{d}x^2}+V(x)+2g|\phi|^2-\mu.
\end{equation}
Complex BdG eigenvalues indicate dynamical (exponential) instabilities, while a purely imaginary spectrum implies linear stability. To further quantify stability, we compute the maximum of the real part of the BdG eigenvalues,
\begin{equation}
	\varepsilon_{\max}\equiv \max \mathrm{Re}\,(\varepsilon),
	\label{eq:epsmax}
\end{equation}
implying that a state is stable if $\varepsilon_{\max}=0$. Our calculations employ a phase-matched box with Fourier collocation (pseudospectral) discretization~\cite{DeconinckKutz2006,WeidemanReddy2000,Yang2008JCP}.

\begin{figure}[t]
 	\centering
 	\includegraphics{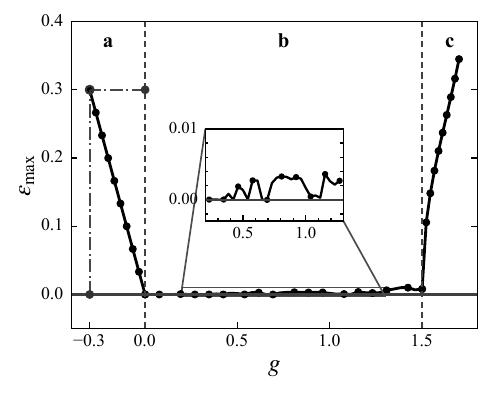} 
    \caption{Stability of the isolated reflectionless state $\phi_{\mathrm{I}}$. $\varepsilon_{\max}$ is plotted versus the interaction strength $g$ at fixed chemical potential $\mu=7/4$. Here $a$, $b$, $c$ denote the three regimes ($g<0$), ($0<g<g_c$), and ($g>g_c$), respectively, separated by two dashed vertical lines. The critical point $g_c=1.5$ is defined by $V_0(g_c)=0$.}
 	\label{fig:fig2}
\end{figure}

For the isolated reflectionless state $\phi_{\mathrm{I}}(x)$ in Eq.~\eqref{eq:phi-FamilyI}, the parameters $V_0$, $\mu$, and $g$ are intrinsically linked by Eq.~\eqref{eq:V0-nl}. To analyze the stability of this solution, we fixed the chemical potential ($\mu=7/4$) and varied the interaction strength $g$. Consequently, the potential depth $V_0$ varies with $g$. Additionally, we require $g < \mu$ to ensure the wavenumber $k = \sqrt{2 (\mu - g)}$ remains real. The stability as a function of $g$ under these conditions is summarized in Fig.~\ref{fig:fig2}. Here, $a$, $b$, $c$ denote three distinct regimes with different stability behaviors. In the focusing regime a ($g<0$), the solution is unstable across the entire range, primarily due to the modulational instability of the homogeneous background. Here, $\varepsilon_{\max} \approx |g|$~\cite{KivsharAgrawal2003,Agrawal2019}.
In the defocusing regime b with $0<g<g_c$, corresponding to an attractive potential well, the solution exhibits instability for most $g$ values but is stable within several narrow ranges. In the defocusing regime c but with $g>g_c$, where the potential well transforms into a barrier, $\varepsilon_{\max}$ grows rapidly with increasing $g$, indicating a stronger instability.

\begin{figure}[t] 
	\centering 
	\includegraphics{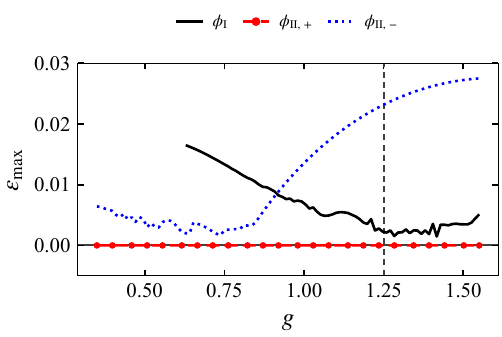} 
	\caption{$\varepsilon_{\max}(g)$ versus the interaction strength $g$ for the reflectionless states $\phi_{\mathrm{II},+}$, $\phi_{\mathrm{II},-}$, and $\phi_{\mathrm{I}}$. The reflectionless states $\phi_{\mathrm{II},\pm}$ exist for $g>1/4$, whereas the reflectionless state $\phi_{\mathrm{I}}$ with fixed $V_0=3/8$ holds only for $g>5/8$ due to $k^2 = 8g /5 - 1$. The vertical dashed line at $g=5/4$ marks the three-fold degenerate point with $V_0=3/8, \mu=7/4, g=5/4$, and $k=1$.} 
	\label{fig:fig3} 
\end{figure}
Figure~\ref{fig:fig3} shows $\varepsilon_{\max}$ as a function of $g$ for the two-fold degenerate reflectionless states. 
For the $\phi_{\mathrm{II},+}$ state, $\varepsilon_{\max}^{(\mathrm{II},+)}(g)=0$ across the entire domain $g>1/4$, indicating stability. In contrast, for the $\phi_{\mathrm{II},-}$ state, $\varepsilon_{\max}^{(\mathrm{II},-)}(g)$ remains finite, revealing instability~\cite{Frantzeskakis2010}. The isolated branch $\phi_{\mathrm{I}}$
exhibits similar instability at fixed $V_0=3/8$. The choice of $V_0=3/8$ results in $\mu(g)=9g/5-1/2$ and $k^2 = 8g /5 - 1$, restricting propagation to $g>5/8$. At the three-fold degenerate parameter point $V_0=3/8, \mu=7/4, g=5/4$, and $k=1$, $\varepsilon_{\max}^{(\mathrm{II},+)}$ vanishes and $\varepsilon_{\max}^{(\mathrm{II},-)}$ and $\varepsilon_{\max}^{(\mathrm{I})}$ remain finite.

Analyzing both families suggests that only one solution per degenerate manifold is dynamically stable. These exact solutions nevertheless elucidate scattering characteristics, implying the existence of additional stable scattering states inaccessible analytically. Thus, our reflectionless solutions provide direct evidence for bistability and multistability structures in this system~\cite{Schlagheck2005a}.

\section{Bistability with Spatially Localized Nonlinearity}
The exact reflectionless states derived in Sec.~\ref{sec:RM-multiple-exact} assume uniform interaction strength. In realistic systems—such as optical propagation in nonlinear media and atoms confined to atomic waveguides—nonlinearity is often spatially localized \cite{Schlagheck2005a,Paul2007,Yamazaki2010}. To explore bistability in such configurations, we analyze stationary states of a BEC with localized nonlinearity via the GPE:
\begin{equation}
	-\frac{1}{2}\,\frac{\mathrm{d}^{2}\phi(x)}{\mathrm{d}x^{2}}
	+ V(x)\,\phi(x)
	+ g(x)\,|\phi(x)|^{2}\phi(x)
	= \mu\,\phi(x),
	\label{eq:embedded-gpe}
\end{equation}
where the interaction strength is modeled by
\begin{equation}
	g(x) = g_0 \operatorname{sech}(x).
\end{equation}
Here $g_0$ is the peak interaction strength, corresponding to finite-range atomic interactions \cite{Dalfovo1999}. Under the parameter conditions
\begin{equation}
    V_0=\frac{5}{8}, \quad \mu=\frac{\lambda^2}{2}-\frac{1}{8}, \quad g_0=\frac{\lambda}{2},\quad k=\sqrt{\lambda^2-\frac{1}{4}},
	\label{eq:embedded-param}
\end{equation}
with $\lambda$ being a real parameter and $|\lambda|>1/2$ ensuring real $k$, the system admits the exact reflectionless state
\begin{equation}
	\phi_e(x)
	= \frac{e^{\mathrm{i} k x}}{\sqrt{e^{-x}+e^{x}}}\,
	\left( e^{-x/2} + \frac{2\lambda}{1+2\mathrm{i} k}\,e^{x/2} \right). 
\end{equation}
This yields unit transmission \(|\tilde{t}|=1\).

\begin{figure}[t]
	\centering
	\includegraphics{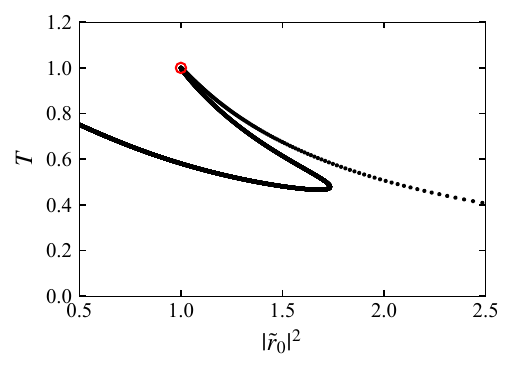}
	\caption{Transmission $T$ versus incident intensity $|\tilde{r}_0|^2$ for the RM potential with a localized nonlinearity $g(x)=g_0 \operatorname{sech}(x)$, calculated from Eq.~(\ref{eq:embedded-gpe}). 
		The red circle marks the analytically derived perfect-transmission state, corresponding to the upper turning point. Here the potential depth is $V_0=5/8$, and other physical parameters are determined by setting $\lambda=3/4$ in Eq.~\eqref{eq:embedded-param}, which yields $ \mu=5/32, g_0=3/8$, and $k=\sqrt{5}/4$.
	}
	\label{fig:fig4}
\end{figure}

To characterize the incident–transmitted intensity relationship, we solved Eq.~(\ref{eq:embedded-gpe}) numerically using the method of Delyon et al.\cite{Delyon1986}. Discretizing in $[-L,L]$ with asymptotic wavenumber $k_1=\sqrt{2 \mu}$, we imposed scattering boundary conditions:
\begin{equation}
	\phi(x)\sim
	\begin{cases}
		\tilde{r}_0 e^{\mathrm{i}k_1 x} + \tilde{r} e^{-\mathrm{i}k_1 x} & x<-L,\\[4pt]
		\tilde{t}\,e^{\mathrm{i}k_1 x} & x>L,
	\end{cases}
\end{equation}
Backward iteration from $x>L$ yields incident amplitude $\tilde{r}_0$ and reflected amplitude $\tilde{r}$, enabling parametric mapping of the transmittance $\left(T \equiv|\tilde{t}|^2 /\left|\tilde{r}_0\right|^2\right)$ against incident intensity $|\tilde{r}_0|^2$.

Figure~\ref{fig:fig4} plots the transmittance $T$ as a function of the incident intensity $\left|\tilde{r}_0\right|^2$. The S-shaped curve exhibits plateaus and folds—hallmarks of bistability—indicating coexisting transmission states over a range of incident intensities. Interestingly, the analytically derived perfect-transmission state (red circle) lies precisely at the upper turning point of the bistable region, confirming the solution and marking the onset of bistability.

\section{Conclusion}
In conclusion, we presented a new family of two-fold degenerate reflectionless states, $\phi_{\mathrm{II},\pm}$, of a BEC in the RM potential, extending a known isolated reflectionless state ($\phi_{\mathrm{I}}$) via a wavefunction ansatz method. Significantly, we uncovered a specific parameter regime where ($\phi_{\mathrm{I}}$) 
coexists with this family, yielding a three-fold degeneracy of perfectly transmitting states characterized by distinct transmission phases. Subsequent BdG analysis of the dynamical stability of these coexisting solutions revealed that only the $\phi_{\mathrm{II},+}$ branch is stable. This result could provide explicit analytical confirmation of the underlying bistable and multistable transmission phenomena.

Furthermore, extending beyond the pure RM potential, we simulated a more realistic spatially localized nonlinearity with a sech profile. Numerically, a prominent bistable transmission characteristic emerged, whose upper critical point is accurately captured by our exact reflectionless state solution. This work unifies analytical methods to demonstrate how nonlinearity fundamentally shapes bistable phenomena. Given the feasibility of realizing RM-like potentials in ultracold atomic gases and engineered photonic lattices, these findings should be directly applicable to such systems. 

\begin{acknowledgments}
This work was supported by the National Natural Science Foundation
of China under Grant No. 11565011.
\end{acknowledgments}

\section*{Declaration of Competing Interest}
The authors declare that they have no known competing financial
interests or personal relationships that could have appeared to influence the work reported in this paper.

\appendix

\section{Derivation of the pairwise-degenerate reflectionless states}
\label{app:family-II}
Here we present a detailed derivation of the pairwise-degenerate reflectionless states $\phi_{\mathrm{II},\pm}$. 
Motivated by the form of the isolated exact reflectionless state $\phi_{\mathrm{I}}$, we take the minimal two-term wavefunction ansatz
\begin{equation}
	\phi(x)=e^{\mathrm{i} k x} f(x)
	=e^{\mathrm{i} k x}\,\frac{\,e^{-x/2}+\tilde{t}\,e^{x/2}\,}{\sqrt{\,e^{-x}+e^{x}\,}},
	\label{eq:app-ansatz}
\end{equation}
which ensures the limits $f(x)\to 1$ as $x\to-\infty$ and $f(x)\to \tilde{t}$ as $x\to+\infty$. Substituting $\phi(x)=f(x)\,e^{\mathrm{i}kx}$ into Eq.~\eqref{eq:stationary} yields
\begin{equation}
	-\frac{1}{2}\,\frac{d^{2}f(x)}{dx^{2}}
	-\mathrm{i}k\,\frac{df(x)}{dx}
	+\Big[V(x)+g\big(|f(x)|^{2}-1\big)\Big]\,f(x)=0.
\end{equation}
Inserting \eqref{eq:app-ansatz}, expressing all terms in exponentials, and clearing denominators by multiplying through by $(e^{-x}+e^{x})^{5/2}$, we obtain a linear combination of $e^{\pm 3x/2}$ and $e^{\pm x/2}$. Setting the coefficients of these exponentials to zero gives the consistency conditions
\begin{equation}
	V_{0}=\frac{3}{8},\qquad k^{2}=g-\frac{1}{4},\qquad
	\label{eq:app-V0}
\end{equation}
and from Eq.~\eqref{eq:dispersion}, $\mu=3g/2-1/8$. Propagating solutions require $k\in\mathbb{R}^+$, implying $g>1/4$.

The solutions for the transmission amplitude $\tilde{t}$ degenerate into a pair of values:
\begin{equation}
	\tilde{t}_{\mathrm{II},\pm}=\pm\,\frac{2\sqrt{g}}{\,1+2\mathrm{i} k\,},
	\label{eq:app-tII}
\end{equation}
whose unimodularity $|\tilde{t}_{\mathrm{II},\pm}|^{2}=4g/(1+4k^{2})=1$ confirms perfect transmission. The solutions differ only by a global phase factor of $\pi$, with $\tilde{t}_{\mathrm{II},-}=-\tilde{t}_{\mathrm{II},+}$.

Substituting \eqref{eq:app-tII} into \eqref{eq:app-ansatz} yields the explicit reflectionless solutions
\begin{equation}
	\phi_{\mathrm{II},\pm}(x)
	=\frac{e^{\mathrm{i} k x}}{\sqrt{\,e^{-x}+e^{x}\,}}
	\left(\,e^{-x/2}\,\pm\,\frac{2\sqrt{g}}{\,1+2\mathrm{i}k\,}\,e^{x/2}\right),
\end{equation}
valid under \eqref{eq:app-V0}. Both branches satisfy the left-incident boundary conditions with $|\tilde{t}_{\mathrm{II},\pm}|=1$.

\section{Derivation of a new family of the pairwise-degenerate reflectionless states}
\label{app:family-III}
Here we show that the trial wavefunction method extends to a new family of pairwise-degenerate reflectionless states. 
We construct analytic reflectionless solutions by augmenting the envelope of the isolated solution with a constant term; the derivation parallels that for $\phi_{\mathrm{II},\pm}$. 
The envelope $f(x)$ takes the form
\begin{equation}
	f(x)=\frac{\,A+e^{-x}+\tilde{t} e^{x}\,}{\,e^{-x}+\,e^{x}\,},
\end{equation}
where $A$ is a constant to be determined. This ansatz manifestly satisfies the scattering limits $f(x)\to 1$ as $x\to-\infty$ and $f(x)\to \tilde{t}$ as $x\to+\infty$.

\begin{figure}[t]
	\centering
	\includegraphics{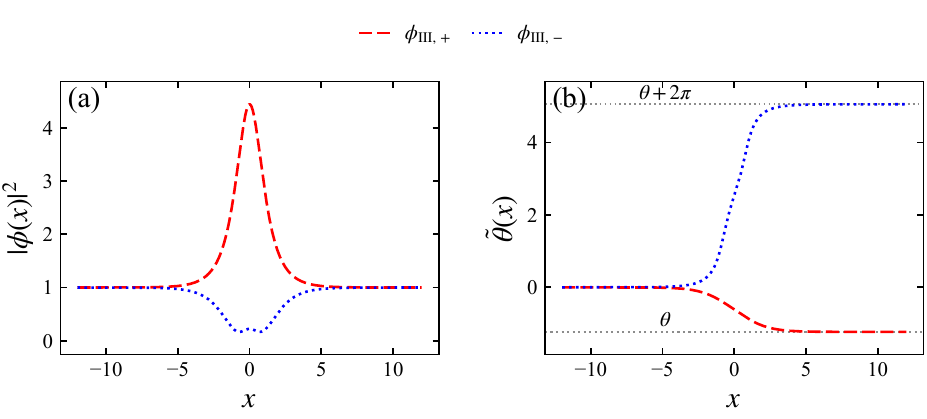}
	\caption{\textbf{(a)} Spatial profiles of the densities $|\phi_{\mathrm{III},+}(x)|^2$ and $|\phi_{\mathrm{III},-}(x)|^2$ of the degenerate reflectionless states $\phi_{\mathrm{III},\pm}(x)$ for the RM well ($V_0 = 3/2$). Parameters: $g = 3/8$ (giving $k = \sqrt{g - 1/4}$ and $\mu = 3g/2 - 1/8$).
		\textbf{(b)} phase $\tilde{\theta}(x)=\arg \left[\phi(x) e^{-\mathrm{i} k x}\right]$; the horizontal line indicates the transmission phase $\arg (\tilde{t}_{\mathrm{III}})$.}
	\label{fig:fig5}
\end{figure}

\begin{figure}[t]
	\centering
	\includegraphics{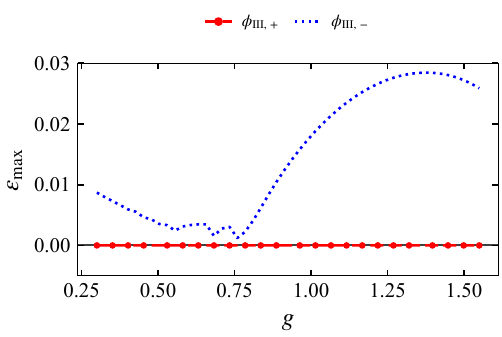}
	\caption{Stability index $\varepsilon_{\max}(g)$ (defined in Eq.~\eqref{eq:epsmax}) versus the interaction strength $g$ for the pairwise-degenerate branches $\phi_{\mathrm{III},+}$ and $\phi_{\mathrm{III},-}$. These branches exist for $g>1/4$ (since $k^2=g-1/4$ under $\mu=3g/2-1/8$ with $V_0=3/2$).}
	\label{fig:fig6}
\end{figure}

Following the procedure for $\phi_{\mathrm{II},\pm}$, we substitute the full wavefunction $\phi(x)=f(x)e^{\mathrm{i} k x}$ into the stationary GPE. Requiring the coefficients of all independent exponentials to vanish yields the algebraic constraints
\begin{equation}
	V_{0}=\frac{3}{2},\qquad
	k^{2}=g-\frac{1}{4}\ \Rightarrow\ \mu=\frac{3}{2}g-\frac{1}{8},\quad g>1/4,
\end{equation}
together with
\begin{equation}
	A_{\pm}=\pm\,\frac{2\sqrt{\,1+4g\,}}{\,1+2\mathrm{i}k\,},\qquad
	\tilde{t}_{\mathrm{III}}=\frac{\,1+4g-2\mathrm{i}k\,}{(1+2\mathrm{i}k)(2+2\mathrm{i}k)}.
\end{equation}
The two sign choices in $A_{\pm}$ define two distinct branches; both share the same transmission amplitude $\tilde{t}_{\mathrm{III}}$ (hence identical transmission phase and $|\tilde{t}_{\mathrm{III}}|=1$).

The resulting pairwise-degenerate reflectionless solutions read
\begin{equation}
	\phi_{\mathrm{III},\pm}(x)
	=\frac{e^{\mathrm{i} k x}}{\,e^{-x}+e^{x}\,}\Big(A_{\pm}+e^{-x}+\tilde{t}_{\mathrm{III}}\,e^{x}\Big),
\end{equation}
and both satisfy the left-incident boundary conditions with unit transmission probability.

Figure~\ref{fig:fig5} shows representative density profiles $|\phi_{\mathrm{III},+}(x)|^{2}$ and $|\phi_{\mathrm{III},-}(x)|^{2}$. While the internal profiles differ through $A_{\pm}$, the far-field transmission phase is identical because their envelope phases differ by exactly $2 \pi$ in the far field, as shown in Fig.~\ref{fig:fig5}(b).

We assess the stability of $\phi_{\mathrm{III},\pm}$ using the BdG methodology described above. Figure~\ref{fig:fig6} shows the stability index $\varepsilon_{\max}(g)$ [Eq.~\eqref{eq:epsmax}] for the two branches under $V_0=3/2$ and $\mu=3g/2-1/8$. Across its entire existence domain, the $\phi_{\mathrm{III},+}$ branch remains dynamically stable, with $\varepsilon_{\max}^{(\mathrm{III},+)}(g)= 0$ within numerical tolerance, whereas the $\phi_{\mathrm{III},-}$ branch exhibits instability, with $\varepsilon_{\max}^{(\mathrm{III},-)}(g)>0$.

\end{document}